# Using Modern Technologies to Capture and Share Indigenous Astronomical Knowledge

Martin Nakata [a]*, Duane Hamacher [a], John Warren [e], Alex Byrne [d], Maurice Pagnucco [b], Ross Harley [c], Srikumar Venugopal [b], Kirsten Thorpe [d], Richard Neville [d] and Reuben Bolt [a]

[a] Nura Gili Indigenous Programs Unit, University of New South Wales
[b] School of Computer Science and Engineering, University of New South Wales
[c] College of Fine Arts, University of New South Wales, Sydney
[d] State Library of New South Wales, Sydney, Australia
[e] Microsoft Research, North Ryde, Australia

**Abstract**

Indigenous Knowledge is important for Indigenous communities across the globe and for the advancement of our general scientific knowledge. In particular, Indigenous astronomical knowledge integrates many aspects of Indigenous Knowledge, including seasonal calendars, navigation, food economics, law, ceremony, and social structure. We aim to develop innovative ways of capturing, managing, and disseminating Indigenous astronomical knowledge for Indigenous communities and the general public for the future. Capturing, managing, and disseminating this knowledge in the digital environment poses a number of challenges, which we aim to address using a collaborative project involving experts in the higher education, library, and industry sectors. Using Microsoft's WorldWide Telescope and Rich Interactive Narratives technologies, we propose to develop software, media design, and archival management solutions to allow Indigenous communities to share their astronomical knowledge with the world on their terms and in a culturally sensitive manner.

**Keywords**: Indigenous Knowledge, cultural astronomy, data management, Intellectual property, Indigenous Australians.

## Introduction

Although increasing numbers of Indigenous people live less traditional lives, many still seek to maintain meaningful connections to their traditional knowledge, heritage, and land (Edwards and Heinrich 2006). Such priorities for the continuance of their traditions and the transmission of Indigenous knowledge to future generations hold a special place in the United Nation's Declaration on the Rights of Indigenous Peoples (United Nations 2008). Major efforts have been made to preserve Indigenous knowledge in accessible forms through recording and documenting traditional knowledge, enabling the retrieval of knowledge in memory and current practice, and identifying and retrieving previously documented knowledge stored in institutions (Nakata and Langton 2005). Vast improvements in digital and virtual technologies now provide a range of new opportunities for many to renew their traditional interests.

Systems of Indigenous knowledge have different verification systems that allow for the renewal of knowledge, specific to the changing conditions of its enactment and practice in situ (Verran 2002). Indigenous people also have different ways of managing access to, and use of, knowledge which is differentiated at three general levels: public areas (open-access), peri-restricted areas (requiring negotiation for access and use terms); and highly restricted or closed areas (secret-sacred knowledge sites, practices and documentation) (Gumbula 2005).

We have all come to understand that Indigenous knowledge[1] derives, produces, and renews its meanings through its practice in situ. It is active, performative, and narrated (not simply expressed through "facts" of it). It is embedded in the local immediate context (not abstracted and generalised across time and place). Its classification processes relate people to place and each other in very different ways to Western understandings (Agrawal 1995, 2002; Christie 2008). The integrity and full meaning of Indigenous knowledge management can therefore be easily overlooked in its translation via the methods and logic of the Western disciplines. The abstraction and re-structuring that occurs in database design can transform and render static dynamic and living knowledge systems, potentially re-shaping the conceptual logic of Indigenous users (Christie 2005).

There are numerous national and international projects aimed at documenting Indigenous knowledge in digital form, and there are many researchers who provide some useful analysis of these practices (e.g., Agrawal 1995, 2002; Ellen et al. 2000; Mathew 2013; Nordin et al. 2012). In a global context, interest in Indigenous knowledge more often forms around its relation to the interests of scientific disciplines, such as environmental science and pharmacology (see e.g., UN Development Program, UN Food and Agriculture Organization, UN Convention on Biological Diversity). Interests within the sciences tend to extract "nuggets" of Indigenous knowledge, detaching them from their social meanings and practices (Eyzaguirre 2001). Development and conservation interests also focus on the value of Indigenous knowledge for practical and general applications elsewhere, storing and reducing knowledge to "data" (Agrawal 1995; Ellen et al. 2000). For these reasons, the scientific and development activity confronts resistance from Indigenous people committed to the protection and on-going use of their cultural and intellectual property interests (Anderson 2005).

In Australia, Yolngu communities in Arnhem Land still utilise traditional (precolonial) knowledge in daily life. Researchers working with these communities (e.g., Christie 2005, 2008; Verran 2005) describe the challenges of databasing Indigenous knowledge in ways that preserve its structural relations. For these communities, digital representation can never be a substitute for traditional practices.

---

[1] See also the International Council for Science's position on Indigenous knowledge "a cumulative body of knowledge, know-how, practices, and representations maintained and developed by peoples with extended histories of interaction with the natural environment. These sophisticated sets of understandings, interpretations and means are part and parcel of a cultural complex that encompasses language, naming and classification systems, resource use practices, ritual, spirituality and worldview" (2002, 3).

However, the need for Indigenous knowledge among urban and dislocated Indigenous communities engages different historical and contemporary conditions, and features a particular desire to reconnect with their knowledge traditions. Digital and virtual technologies cannot overcome the distortions to Indigenous knowledge that come with its abstraction into ex situ situations. But they can offer a means for addressing the reductive tendencies of abstracted/extracted knowledge by presenting a more complex array of representations and, thus, a fuller account of its social meanings and significance, as well as enabling dialogic engagement with living bodies of knowledge on traditions. Therefore, it is critically important to utilise technologies with the appropriate capabilities to manage the circulation of Indigenous knowledge in a culturally sensitive manner, and to ensure Indigenous people's on-going engagement with their knowledge traditions.

In Australia, the provision of online engagements with Indigenous people's knowledge is being attempted, but it is happening in a haphazard way due to a range of constraints (Nakata et al. 2008). These online engagements range from restricted online databases, to stand-alone databases with community-only access, to public webpages of varying standards (Scott 2004). Given the increasing convergence between Indigenous community interests in scattered public domain Indigenous knowledge materials and those of wider communities of interest, the quest to evolve a more sustainable model for demonstrating and guiding important culturally distinct knowledge for future activity is on-going. A key problem has been how to re-present this knowledge in ways that denote its local context, its historical journey through archives and the disciplines, and its onward use and engagements according to Indigenous protocols of knowledge management. There is arguably still much to be gained by shifting our focus more squarely on the production process of making and re-making Indigenous knowledge.

**Exploring Indigenous Astronomical Knowledge**

Much of the social organisation and daily practices of Indigenous communities are based on their astronomical knowledge (Johnson, 1998). This includes using the sun, moon, and stars for predictive purposes to guide navigation, time keeping, seasonal calendars, hunting, fishing, and gathering (Hamacher and Norris, 2011). The stars also inform sacred law, customs, and social structure, such as ceremony, totem and kinship classes, and marriage systems. Astronomical knowledge is generated by observing the changing positions of stars, the rising and setting position of the Sun on the horizon, and the monthly phases of the Moon (Hamacher, 2012; Sharp, 1993). Astronomical knowledge is then transmitted to successive generations through oral and material traditions, including story, song, dance, artefacts, rock art, stone arrangements, ceremony, and everyday social practices (Johnson, 1998; Cairns and Harney, 2004). In turn, these traditions describe the origins of the landscape, plants, animals, and people, along with their relationship to each other (Haynes, 2000). Indigenous astronomical knowledge links many aspects of traditional culture with the physical environment, and integrates Indigenous worldviews, philosophies, knowledge organisation, and social traditions. By engaging the many different aspects of Indigenous astronomical knowledge, we are provided with insights into the complexity and diversity of Indigenous knowledge systems. The study of Indigenous astronomical knowledge falls under the academic field of *cultural astronomy*—the study of social understandings and applications of astronomical knowledge (e.g.

Holbrook et al., 2008). For the reasons discussed above, Indigenous cultural astronomy is an excellent area in which to trial and devise solutions to collect, manage, and disseminate a diverse range knowledge.

**Developing Models for Digitally Capturing and Disseminating Indigenous Astronomical Knowledge**

Digital technologies offer avenues for preserving Indigenous astronomical knowledge and making it accessible to future generations of Indigenous people. These technologies also allow for the discovery and retrieval of documented astronomical knowledge held in various collecting institutions, such as library and museum archives. Since Indigenous astronomical practices (like all forms of Indigenous Knowledge) are dynamic, they are constantly being renewed. Therefore, any attempt to manage, represent, or restore this knowledge requires ongoing collaboration between Indigenous knowledge holders, collecting institutions, and researchers.

To manage and disseminate Indigenous astronomical knowledge, one must explore how best to capture and collect Indigenous astronomical knowledge in consultation with the owners of the knowledge while protecting its cultural, practical, and social significance. One must also explore ways of re-examining and reinterpreting Indigenous astronomical knowledge collected by early anthropologists and missionaries that are culturally sensitive and that recognise the contexts in which the knowledge was collected. To accomplish this, we developed a collaborative project involving researchers, Indigenous scholars, archivists, and industry partners with relevant expertise, knowledge, and experience from the university, library, and industry sectors. These organisations are:

a) **The University of New South Wales.**

   Academics in the Nura Gili Indigenous Programs Unit bring expertise in Indigenous studies, community engagement, and cultural astronomy. Indigenous academics will assist with addressing the needs and interests of Indigenous end-users. An academic with expertise in both astrophysics and Indigenous cultural astronomy will help guide the type of information being captured, with assistance from Indigenous team members on all aspects involving Indigenous community engagements.

   Academics in the School of Computer Science & Engineering (CSE) and the College of Fine Arts (CoFA) will bring expertise in artificial intelligence, particularly in the areas of knowledge representation, reasoning, belief change, action and cognitive robotics. This will shape the overall design of the online interface and repository functions. This is supported by additional expertise in cloud-based platforms, data organisation in distributed systems, and mechanisms for large-scale use of cloud elasticity. Software programmers will adapt open-source software to capture and store Indigenous astronomical knowledge.

   The College of Fine Arts and the UNSW iCinema Centre for Interactive Cinema Research will also provide expertise in creative practices and digital/media arts. They will facilitate the design of the presentation platform to ensure functionality and ease-of-use to users. They will produce creative ways of combining images, film, performances, sound, and artwork.

b) **State Library of New South Wales.**

Librarians and archivists at the State Library bring expert knowledge and experience in the contentious areas of intellectual property, cultural sensitivities, and the mechanisms required to ensure best standards of practice in the presentation of digitised Indigenous materials for access by the public. Library team members also have extensive community experience in managing and opening Indigenous knowledge archives with the right cultural protocols.

c) **Microsoft Research**

Microsoft industry partners will help facilitate and guide engagements regarding back-end programming, including the WorldWide Telescope, Rich Interactive Narratives, and Azure (which will be discussed below). The combined team covers all of the areas of expertise that are required to make this project come to fruition.

**Implementing the Project**

A key aspect to our thinking on these complex issues at the interface of Indigenous knowledge and Western systems is in the building of a metadata framework and an overall ontological platform that can facilitate the Indigenous community's initial access points, the capturing of associated information on each contribution of astronomical knowledge and, in ways similar to Wikis, their ongoing sharing, engagements and renewal. We see three tasks ahead of us:

a) How we can best capture and collect Indigenous astronomical knowledge for the future utility of Indigenous communities while keeping its significance for Indigenous cultural, practical, and social organisation intact?
b) How this knowledge-capture should be structured as part of the world's Indigenous heritage, and as part of the growing field of cultural astronomy?
c) How best ways to present and visualise this knowledge in an open and meaningful way for both Indigenous communities and the wider international community?

The approach being developed for this emerging project attempts technical and theoretical methods that avoid representing this "socially-moored" knowledge as a fixed, discrete, or complete body of knowledge that is antiquated in its relationship to Western astronomy. Any representation of Indigenous astronomical knowledge in the digital domain must be able to manage the conversations and interactions produced through online community and scholarly interactions.

We will seek to do this by using two key tools: Microsoft's WorldWide Telescope[2]

---

[2] http://www.worldwidetelescope.org

and Rich Interactive Narratives[3] (RIN). The WorldWide Telescope (WWT) is a free, popular and innovative global community resource, developed by Microsoft Research with more than 10 million users. It enables viewers to explore the night sky by panning and zooming at different scales to reveal different astronomical objects. It also simulates the horizon, allowing the user to see the rising and setting of celestial objects at different times of the day throughout the year, which is essential for understanding Indigenous astronomical knowledge. Using the latest research from astrophysics and cosmology, the WWT is able to retrieve images and data and blend them seamlessly to create an immersive experience. RIN, also developed by Microsoft Research, has new visualization technologies that will help combine traditional forms of storytelling to create compelling interactive digital narratives. There are no equivalents to these tools in terms of their features and capabilities, making them crucial to achieving the aims of this project. The representation and reasoning functions will link Indigenous content and media to astronomical objects and coordinates in WWT, and to narratives in RIN.

**Developing a system for the collection of Indigenous astronomical knowledge and structuring the collection to provide insight into Indigenous knowledge systems**

It is expected that the data collection process will produce media in various forms, such as photographs, documents, videos and webpages. Since the collection will be an on-going process, the virtual repository must store and accommodate an increasing volume of data throughout its lifetime. The media artefacts will be linked to scientific astronomical objects and concepts, such as planets, stars, and star groups. This requires the virtual repository to host an ever-growing collection of media in various formats that are indexed as they are entered so that they can be quickly retrieved by the presentation platforms. From a computer science perspective, this poses a number of challenges. This includes storing unbounded amounts of data for quick retrieval; representing data for both Indigenous astronomical knowledge and astrophysics; and logical reasoning about the relationships between user-generated media objects (which are primarily grounded in Indigenous knowledge and worldview) and astronomical objects that are organised according to Western concepts (e.g., astrometric catalogues, constellations, etc.).

The final design of the system will leverage cloud-computing features to accommodate the ever-growing collection of media. Microsoft's Windows Azure is presently one of the most popular cloud services in the world and provides the computing and storage resources, and the mechanisms to develop services that can draw upon them. Microsoft will provide access to Azure and its many capabilities, while the content will also be archived by the SLNSW for ensuring the longevity of the collection beyond the lifetime of the project. The proposed data capture, storage, and retrieval system will be developed around some of the key features offered by Azure in ways that require little or no Indigenous community resources in order to participate. The project will also look toward developing a decentralised indexing system to facilitate the organisation and quick retrieval of data from the storage system. The Web Ontology Language[4] (OWL), and its associated tools and reasoners, provides one possible approach. OWL currently assumes a Western taxonomical and hierarchical approach to the classification and structure of knowledge. This needs to be re-evaluated to determine a structured knowledge representation that will be best suited to Indigenous knowledge systems. The guidance of our Indigenous experts on the team will be crucial in this process.

---

[3] http://research.microsoft.com/en-us/projects/rin/
[4] http://www.w3.org/TR/owl-features/

**Developing a platform for the presentation of the acquired Indigenous artefacts and knowledge**

This task will involve the development of a platform that retrieves content from the storage system and presents it in an intuitive manner so that users can discover links between astronomical objects, their Indigenous interpretations, and astrophysical properties. A significant advantage of using the WWT over other astronomical programmes is a presentation environment that demonstrates the relative positions of the stars with respect to each other and the landscape. Indigenous astronomical traditions frequently reference the rising or setting time and positions of particular stars on the horizon. The WWT provides an extremely accurate environment in which to simulate this. This allows the user to see the positions of celestial objects from any place on Earth, at any time, and from any year. WWT also takes into account orbital effects, stellar proper motion, and other positional astronomy data, which is essential for accurately observing the sky over long periods of time. The computer science members of the team will develop software to "glue" WWT to the proposed data storage system. This will allow WWT to quickly retrieve all user-generated media of relevance to a particular celestial object, or at the confluence of the sky and the land (e.g., rising or setting positions of celestial bodies). With these capabilities, Indigenous communities will be able to develop their own style of narratives and tours in WWT to communicate their astronomical knowledge to the world.

**On-going user experience evaluations of the platform**

The long-term success of the online facilities talked about for this project depends on its adoption by Indigenous communities as a low-cost and user-friendly option to capture and renew their astronomical knowledge. This will involve testing the proposed platform for usability and acceptance. Our first goal will be to customise both the collection and the presentation interfaces to suit the needs of Indigenous communities. The collection interface should, at the end of the day, enable users to upload their media as seamlessly as possible, keeping in mind that some Indigenous communities are located in remote areas with bandwidth issues. The presentation interface needs to be intuitive and allow the appropriate access rights that adhere to cultural sensitivities and protocols. We will also seek to test key risk management strategies using experience from past projects in the library and archive sector (e.g., Nakata et al. 2008). A staged "trial-developimprove" agenda is key to our approach. It will begin with local engagements before moving to remote, regional, national, and international settings. We will need to develop simple interface features and test them initially with two distinctly different Australian Indigenous communities (remote/metropolitan, Aboriginal/Islander dimensions). This will enable us to make improvements early on before moving on to broader applications in other regions. Monitoring the end-user experience with the developing facilities will be an on-going task. Evaluations (Hofstede 2005) and evaluation reviews (George et al. 2012) of cultural dimensions in interface designs will help us develop an appropriate survey tool for the project. Quantitative analytical and survey tools, such as QUIS[5], will be adapted for the project's purposes.

**The Importance of Collaboration**

The mistreatment of Indigenous knowledge has led to reluctance among many Indigenous people to allow their traditional knowledge to be accessible in the public domain (Fourmile 1989). There remains a range of problems in representing Indigenous knowledge, particularly at the intersection of Indigenous Knowledge with Western forms of production, organisation, and management. At the same time, there is resentment among Indigenous people that their traditional knowledge is not acknowledged or valued as a useful contribution to environmental practices, land resource management, or broader scientific knowledge. Our test-site approach enables cultural, archival and technological collaborators to work across a range of problems encountered in the re-presentation of Indigenous knowledge materials at the interface where Indigenous and Western knowledge production, organisation and management collide.

Microsoft Research has a continuing interest in understanding, facilitating, and increasing the global community's engagement with the world's collective astronomical knowledge, not just the interests of the scientific community. Early collaborations between Microsoft Research and the UNSW Australia researchers revealed the extent of the challenges involved in capturing Indigenous knowledge and in increasing Indigenous-user interactions. These challenges encouraged Microsoft to develop the capacity of the WWT to respond to Indigenous needs and interests as part of its commitment to enabling global access and participation. Although this project is, in part, a test site for solving problems for Microsoft Research, it is anticipated that the innovation and knowledge gained from this collaborative work will lead to a more significant representation of the astronomical knowledge of all cultures through the WWT.

We envision that advances and innovations in technological design will stimulate Indigenous community interest, as well as national archival and academic interest in recording and preserving Indigenous astronomical knowledge. This is significant for the preservation of Indigenous heritage worldwide, which is recognised as a foundation for the heritage of all nations. By eventually expanding this project globally, the emerging collaboration will advance technological knowledge and user management capacities for the representation of both Indigenous astronomical knowledge production and Western astronomy.

The involvement of the State Library of New South Wales (SLNSW) will bring focus in the Australian library and archive sector regarding the future management of Indigenous collections. Since it is a government supported priority for the SLNSW to digitise its collections, the SLNSW is committed to developing, trialling, and documenting the best standards of practice for the digitisation of Indigenous Australian materials for the public domain. Australia has been an international leader in Indigenous knowledge and information management protocols for two decades, contributing at the United Nations level and to developments in professional contexts across the globe (Byrne et al. 1995). This project seeks to build upon these strengths and contribute to the growing leadership of Australia in this area.

---

[5] http://lap.umd.edu/quis/QuantQUIS.htm

**Conclusion**

Indigenous astronomical traditions are underpinned by a philosophy of knowledge that enables a different view of the place in which humans relate to the natural world. Indigenous knowledge holds potential for understanding a range of environments in conjunction with knowledge of the stars and planetary cycles. This emerging project seeks ways to improve Indigenous access, utilisation and engagement with documented and undocumented cultural knowledge, as well as ways that give recognition and value to Indigenous knowledge within public repositories that contain predominantly scientific knowledge. The restoration and maintenance of Indigenous knowledge is recognised as a vital ingredient for restoring and maintaining the health and wellbeing of Indigenous communities worldwide. The diversity of human knowledge of the skies, the utilisation of this knowledge by different societies, and the advances of knowledge that comes from scientific endeavour and inquiry will help build a more complete picture of human engagement with the meaning of the stars, which have been on view since time immemorial.